\documentclass[aps,prl,preprint,superscriptaddress,showpacs]{revtex4}

\include{epsf}

\begin{document}


\title{Measurement of the Casimir force between parallel metallic
surfaces}



\author{G. Bressi}
\affiliation{INFN, Sezione di Pavia, Via Bassi 6, Pavia, Italy 27100}
\author{G. Carugno}
\affiliation{INFN, Sezione di Padova, Via Marzolo 8, Padova, Italy 35131}
\author{R. Onofrio}
\affiliation{INFN, Sezione di Padova, Via Marzolo 8, Padova, Italy 35131}
\affiliation{Dipartimento di Fisica 'G. Galilei', Universit\`a di
Padova, Via Marzolo 8, Padova, Italy 35131}
\author{G. Ruoso}
\affiliation{INFN, Sezione di Padova, Via Marzolo 8, Padova, Italy 35131}
\affiliation{Dipartimento di Fisica 'G. Galilei', Universit\`a di
Padova, Via Marzolo 8, Padova, Italy 35131}
\thanks{Contact Information: giuseppe.ruoso@lnl.infn.it}
\email[]{giuseppe.ruoso@lnl.infn.it}
\date{\today}

\begin{abstract}
We report on the measurement of the Casimir force between conducting 
surfaces in a parallel configuration. 
The force is exerted between a silicon cantilever coated with chromium 
and a similar rigid surface and is detected looking at the shifts induced 
in the cantilever frequency when the latter is approached. 
The scaling of the force with the distance between the surfaces was tested 
in the 0.5 - 3.0 $\mu$m range, and the related force coefficient was 
determined at the 15\% precision level.
\end{abstract}

\maketitle

One of the most astonishing features of quantum physics is that, as an 
ultimate consequence of the uncertainty principle, the vacuum is not empty.
The nontrivial structure of the quantum vacuum has profound implications
at both the microscopic and macroscopic level. In particular, forces of
extra-gravitational origin acting between neutral bodies have been predicted
due to the deformation of vacuum fluctuations caused by the macroscopic
boundary conditions. In recent years, a compelling motivation to better 
grasp the contribution of quantum vacuum to the space-time curvature 
\cite{Zeldovich,Weinberg,Witten} is provided by the reported evidence 
for an accelerating Universe \cite{Riess,Garnavich,Perlmutter}.
A first-principle calculation of the  pressure due to  zero-point 
electromagnetic fluctuations for the case of an indefinite plane cavity 
made of conducting materials spaced by a distance $d$ is obtained by 
summing all the vacuum modes contributions \cite{Casimir}. 
This results in the prediction of the quantum vacuum pressure as 
$P_C=K_C /d^4$, where  the coefficient 
$K_C=\pi h c/480=1.3 \cdot 10^{-27}$ N m${}^2$ has been introduced, 
with  $h$ and $c$ 
denoting the Planck constant and the speed of light, respectively.

Several experimental attempts have been pursued during various decades for
unambiguously verifying Casimir's prediction \cite{Bordag}.     
So far, this search has been successful only in a particular geometry, namely 
in a cavity constituted by a plane surface opposing a spherical one. 
Pioneering measurements by van Blokland and Overbeek \cite{Vanblokland} 
in such a configuration resulted in the observation of the associated Casimir 
force, and in its detailed comparison to the Lifshits theory \cite{Lifshits} 
taking into account finite conductivity effects.
More recently, these measurements have been revived by using state of art 
torsion balances \cite{Lamoreaux}, atomic force microscopes \cite{Mohideen}, 
and high precision capacitance bridges \cite{Chan}. The latter two 
experiments have reached 1 \% precision, more precise determinations 
being limited by the theoretical uncertainty due to the so-called 
proximity force theorem (see section 4.3 in \cite{Bordag} for details).
Regarding the Casimir force between two parallel conducting surfaces, the 
situation originally discussed by Casimir, no clear experimental result has 
been obtained so far. The only  attempt in this configuration 
dates back to Sparnaay \cite{Sparnaay}. 
The experimental data he obtained ``do not contradict Casimir's theoretical 
prediction'' \cite{Sparnaay}, but large systematic errors and uncontrollable 
electrostatic forces prevented a detailed quantitative study.
In this Letter we report on the measurement of the Casimir force between 
parallel conducting surfaces in the $0.5 - 3.0 \,\,\mu$m range with 
15 \% precision. 
Our results are expected to have far-reaching implications toward
understanding the nature and role of  quantum  fluctuations at the
macroscale, as well as for exploring gravity at the microscale.

In our experiment, the two parallel surfaces are the opposing faces 
of a cantilever beam, free to oscillate around its clamping point, 
and of another thicker beam rigidly connected to a frame with 
adjustable distance from the cantilever. 
Our apparatus has already been discussed in detail elsewhere \cite{Bressi}, 
and a schematic of the experimental set-up is shown in Fig. \ref{set-up}.
We use a rectangularly shaped cantilever made of  silicon
(resistivity 10 $\Omega $ cm), with optically flat surfaces 
of size 1.9 cm $\times$ 1.2  mm $\times$ 47 $\mu$m (average roughness 
$\sim 10$ nm), covered with a 50 nm thick chromium layer.
The resonator is clamped to a copper base by which it can be rotated around 
the horizontal axis, parallel to its  faces, by using a nanometer step motor. 
The resonator is faced on one side by another silicon beam (hereafter called 
the source), placed along the orthogonal direction and also covered by 
a (thicker) chromium layer.  This beam has the same longitudinal dimensions 
 of the first one (1.9 cm $\times$ 1.2  mm) but is much thicker (0.5 mm). 
The source beam can be rotated by using step motors around the two axes 
complementary to the one controlled by the resonator tilting, thus 
providing a fine control  of the parallelism of the two opposing surfaces. 
The gap separation between the two surfaces is adjusted with a dc motor 
for the coarse movement, and finely tuned using a linear piezoelectric 
(PZT) ceramic attached to the source. 
The source and the resonator are electrically connected to a voltage 
calibrator for the electrostatic calibrations or, alternatively, to 
an ac bridge for measuring the capacitance of the system.
We detect the motion of the resonator by means of a fiber optic 
interferometer \cite{Rugar} located on the opposite side of the resonator.
The interferometer detects the relative displacement between the resonator and 
the detection fiber end, with a typical sensitivity of $1.0\times10^{-7}$ m/V.
The lowest torsional mode of the cantilever is monitored, its 
free frequency being $\nu_0=138.275$ Hz and the mechanical quality factor 
$\sim 10^3$.
The major problems, common to previous experimental efforts, are attributable 
to the difficulty to achieve and control the ideal conditions of parallel and 
neutral surfaces. The two surfaces must be kept parallel even at the smallest 
gaps investigated. Also, due to the presence of different metals in the 
electrical circuit connecting the two surfaces, an offset voltage $V_{0}$ is 
always present in the gap, even when the two surfaces are nominally 
short-circuited. This voltage prevents the possibility to obtain small gap 
separations because the electrostatic force will cause the resonator to 
attach on the source. For these reasons the measuring process can be divided 
into three stages: parallelization of the gap between the two 
surfaces, 
on-line estimate of the offset voltage $V_{0}$, and 
calibration with electrostatic fields, including one canceling 
the effect of $V_{0}$ at the leading order. This last stage 
allows to reach the small separations ($\leq 1 \mu$m) 
at which the Casimir force, thanks to its favourable scaling, is expected 
to dominate over the residual electrostatic force. 

A prerequisite for the parallelization procedure is the stabilization and the 
elimination of dust particles present on the two surfaces. 
A SiO$_{2}$ surface etching and the evaporation of a deposit of chromium 
provide stable metallic surfaces, and also prevents a fast oxydation.
For the cleaning, besides adopting standard procedures as a dust-free laminar 
air flow environment able to filter powders of less than 1 $\mu$m, and 
washing with proper solvents, we use a dedicated in-vacuum cleaning tool 
\cite{Bressi}. 
The latter is made of a thin metallic wire which, under inspection at the 
SEM, is moved along three orthogonal axes through vacuum feedthroughs with 
micrometers. Dust grains of size between 0.5 and 3 $\mu$m, difficult to 
identify under the optical microscope used for the preliminary, in air, 
cleaning, are then removed. Once the surfaces are cleaned at the 
0.5 $\mu$m level, we optimize their parallelism. A coarse arrangement is 
first done by using the SEM, i.e. viewing the gap at different magnifications 
on two orthogonal sides (see for instance Fig. \ref{sempicture}). 
By means of the various motion controls it is possible 
to reach an almost parallel configuration (within a 1 $\mu$m resolution). 
The final parallelism is then obtained using the ac bridge by 
maximizing the capacitance at the minimum obtainable gap separation. 
A maximum value of 22 pF is obtained, corresponding to an average 
gap separation of about 0.4 $\mu$m. With an ac bridge sensitivity of  
$\sim 0.4$ pF and based on the expression for the capacitance between non 
parallel plates this guarantees a parallelism better than 30 nm over 1.2 mm, 
equivalent to an angular deviation of $\sim 3 \cdot 10^{-5}$ radians.

To control deviations from electrical neutrality of the two surfaces and 
to obtain a rough on-line determination of $V_{0}$, we measure the static 
deflection of the resonator versus the external voltage $V_{c}$ applied 
with the calibrator for various gap distances $d_{i}$. 
The bending is measured by looking at the dc level of the fiber optic 
interferometer signal, and a repetitive procedure (by alternating 
bias voltages and zero voltage measurements) is adopted to cancel out 
the effect of drifts in the laser frequency. 
The static displacement $\Delta x_i(V_{c})$ of the resonator at its top edge 
is given by $\Delta x_i(V_{c})=K_i (V_{c}-V_{0})^2$.  
Fitting the measured data for each distance $d_{i}$ with this law we obtain 
an average value of $V_{0} = -(68.6\pm 2.2)$ mV. 
The parameter $K_i=\epsilon_0 S/8\pi^2 m_{\rm eff} \nu_0^2 d_i^2$ 
can also be used to evaluate the effective mass of the torsional mode, 
which is $m_{\rm eff}=(0.30 \pm 0.05) \,m_0$, with $m_0$ the physical mass, 
in agreement with theory \cite{Sader}.

By canceling the leading contribution of the offset voltage $V_{0}$ through a 
counterbias voltage $V_{c}\simeq V_{0}$, it is possible to look for 
distance-dependent forces superimposed to the residual bias $V_r=V_{c}-V_0$. 
This is done using a dynamical technique, {\it i.e.} by measuring the resonant 
frequency of the cantilever $\nu$ as the gap separation 
is reduced. Any spatially-dependent force is expected to induce a frequency 
shift whose sign is dependent on the attractive or repulsive nature of the 
force, negative frequency shifts signaling the presence of attractive forces. 
For a superposition of a residual electrostatic force contribution and the 
expected Casimir force the frequency shift is expressed as \cite{Bressi}:
\vspace{-.2cm}
\begin{equation}
\label{eq:sh}
\Delta \nu^{2}(d)= \nu^{2}-\nu_{0}^{2} =
-C_{\rm el}{ V^{2}_{r}\over d^3}-{C_{\rm Cas} \over  d^5},
\end{equation}
where $C_{\rm el}=\epsilon_0 S/4 \pi^2 m_{\rm eff}$ and 
$C_{\rm Cas}=K_C S/\pi^2 m_{\rm eff}$, with $\epsilon_{0}$ the vacuum 
dielectric constant, $S$ the effective area delimited by the cantilever and 
the source surfaces, and $m_{\rm eff}$ the effective mass of the resonator 
mode \cite{Sader}. 

In order to disentangle the two contributions in the right hand side of 
Eq. (\ref{eq:sh}), the measurements are performed in four situations 
differing by the applied bias voltage $V_{c}$.
A delicate issue is the determination of the distance: all the measurements 
with variable gap are done keeping the resonator fixed and moving 
the source by means of the linear ceramic PZT, always using increasing 
voltages to avoid hysteresis effects. The relative displacement 
between the source and the resonator is then expressed, in terms of 
the voltage $V_{\rm PZT}$ applied to the linear PZT, by
$d_{r}=d_{r}^{0}-A V_{\rm PZT}-d_s(V_{\rm PZT})$, where
we have introduced the distance corresponding to $V_{\rm PZT}=0$ V as  
$d_{r}^{0}= 1.2 \cdot 10^{-5}$ m, the actuation coefficient 
$A=(1.508 \pm 0.002) \times10^{-7}$ m/V, and $d_{s}$ which takes 
into account the static deflection of the resonator due to the force. 
These last two quantities are measured with the fiber optic 
interferometer. The precision on the determination of $d_r^{0}$ is very 
critical and we could not rely for its evaluation on the direct 
measurement at the SEM alone. For this reason the electrostatic 
calibration is also used to determine the correcting parameter $d_{0}$, 
such that $d=d_r+d_0$ is the actual gap separation.
Three out of the four measurements are done at large values of the 
bias voltage $(V_{c}=[-205.8,-137.2,+68.6\,\, {\rm mV}])$ and 
large distances. 
Fig. \ref{calibration} shows $\Delta \nu^{2}$ versus the relative 
displacement $d_{r}$. The data are fitted with the function:
\vspace{-.2cm}
\begin{equation}
\Delta \nu^2(d_r)= - \Delta\nu^2_{\rm offset} - 
C_{\rm el}\frac{ V_{r}^{2}}{(d_r+d_0)^{3}},
	\label{eq:fit}
\end{equation}
where $\Delta\nu^2_{\rm offset}$ is a free frequency offset taking into 
account long term drifts. From a global fit we obtained the values 
$\Delta\nu^2_{\rm offset}=(6\pm 1)$ Hz$^2$,  
$d_{0}=-(3.30 \pm 0.32)\cdot 10^{-7}$ m,
$C_{\rm el}= (4.24 \pm 0.11) \cdot 10^{-13}$ Hz${}^2$ m${}^3$, 
and $V_{0}=(60.2\pm 1.7)$ mV, 
with a $\chi^{2}$ probability of 85\%. 

\begin{figure}[htb]
\caption{\label{casimir}
Observation of the Casimir force. 
Residuals of the square frequency shift versus the gap distance and best 
fit with Eq. (\ref{eq:cas}). The fit is done by considering 
the points at the 9 smallest distances (0.5 - 1.1 $\mu$m region), 
and includes the estimated errors, coming from the parameters $C_{\rm 
el}$, $d_{0}$ and $V_{0}$, both for the frequency shift and for the gap 
distance. The use of the 9 points showing the largest shifts comes from a 
$\chi^{2}$ analysis. With this choice the resulting $\chi^{2}$ probability 
is 61\%. A global fit also including a square frequency shift linearly 
increasing in time, with shift values ranging from 0 to 50 Hz$^{2}$, 
allows for a possible explanation of the anomalous frequency shift evident 
in the $ 1 - 2 \,\mu$m region. This gives a force coefficient 
$K_C=(1.24 \pm 0.10) \cdot 10^{-27}$N m${}^2$, almost identical to the one 
previously found and with a $\chi^{2}$ probability of 55\%. 
Leaving the exponent of $d$ as a free parameter leads to a best fit with 
exponent 5.0 $\pm$ 0.1 as expected from the dynamical 
component of the Casimir force between parallel surfaces.}
\end{figure}
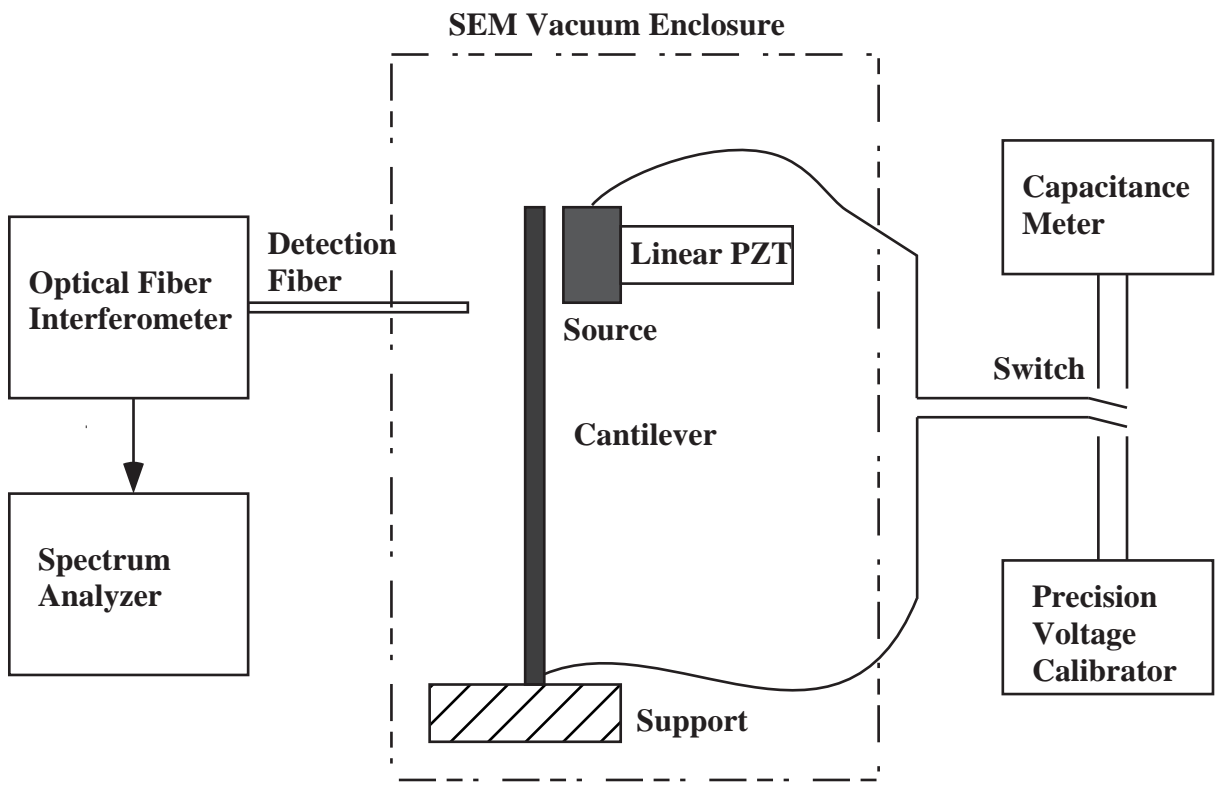
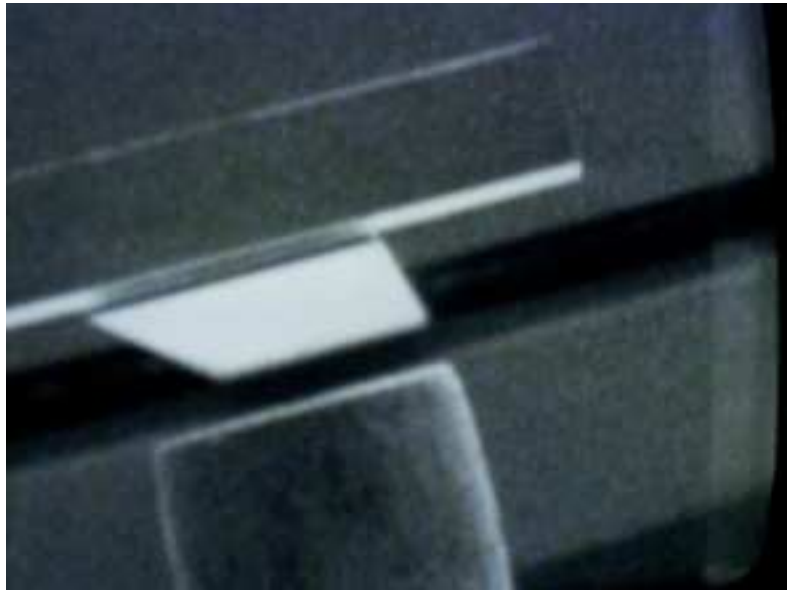
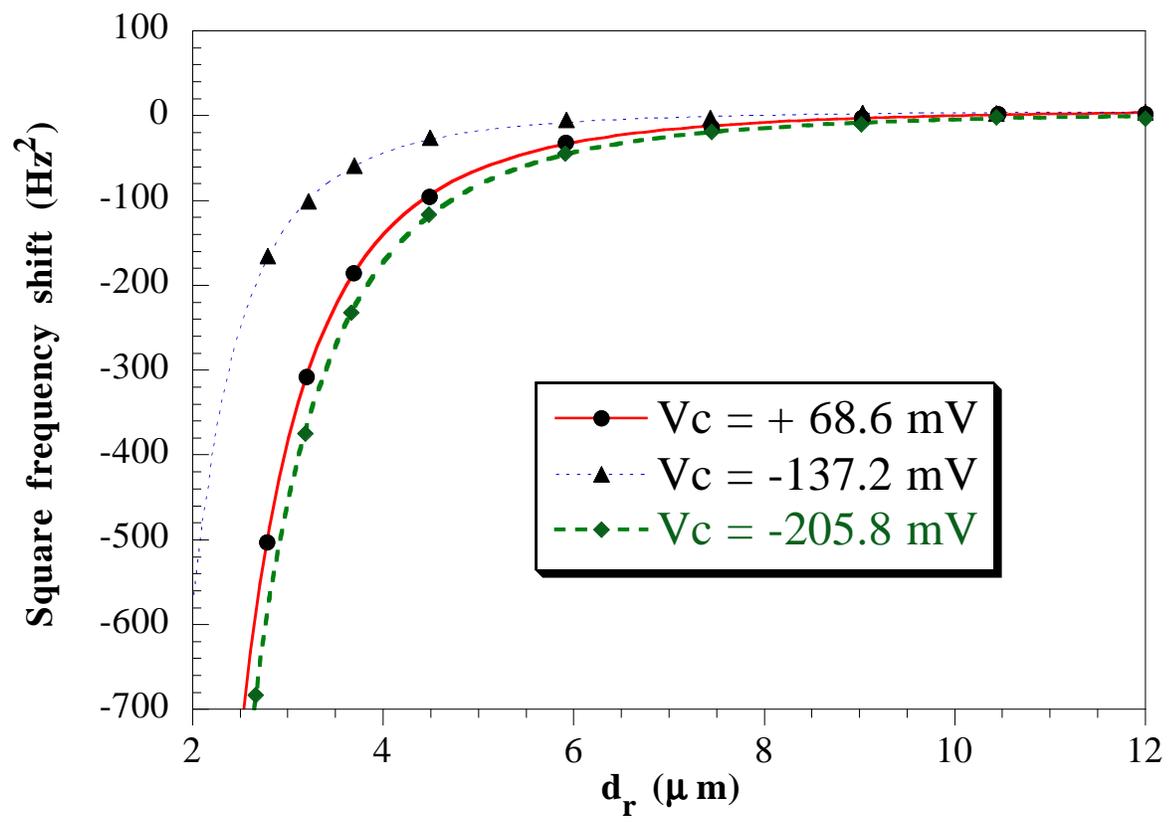
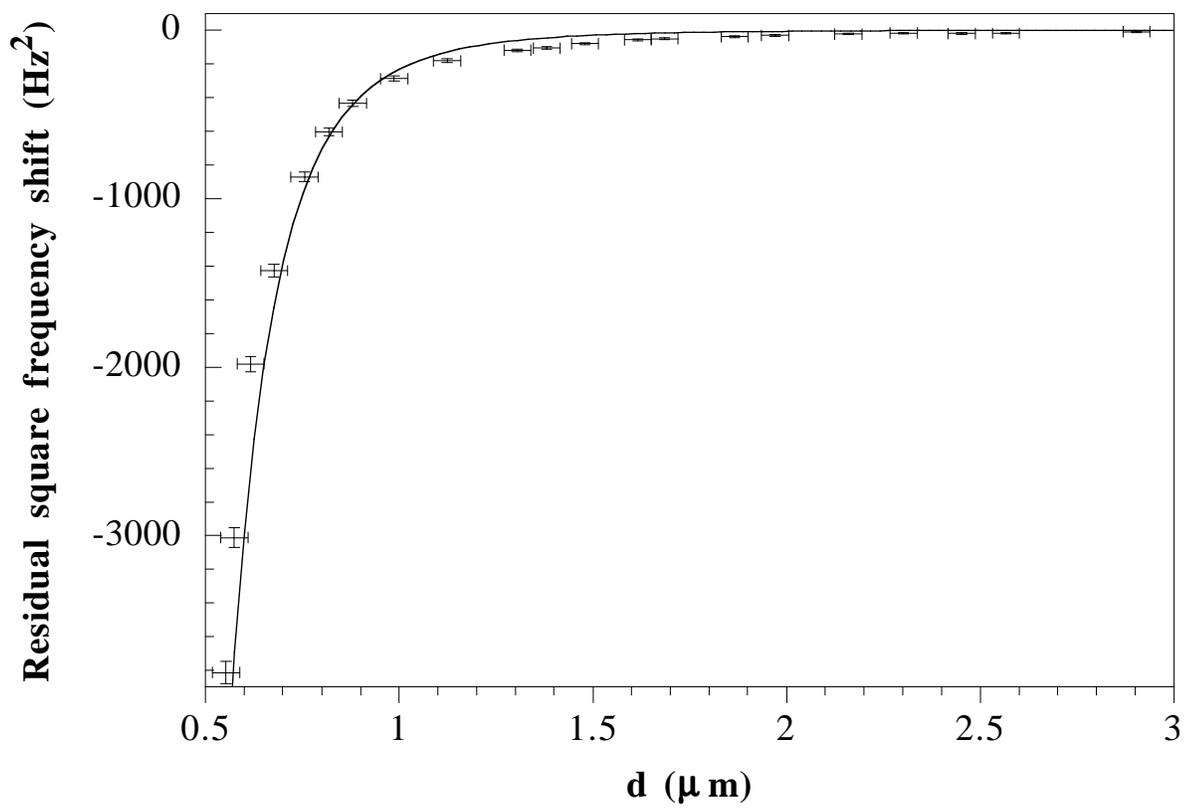

It is now possible to analyze the data set for the fourth situation of 
bias voltage $V_{c}=-68.6$ mV, corresponding to a quasi complete cancellation 
of the effect of the offset voltage. From the acquired knowledge of the 
electrostatic component of the force we subtract its contribution and look 
at the residual frequency shift. The result is shown in Fig. \ref{casimir} 
together with the best fit with the function: 
\vspace{-.2cm}
\begin{equation}
\Delta\nu^2(d)= - \frac{C_{\rm Cas}}{d^{5}}.
\label{eq:cas}
\end{equation} 
This results in a value for the parameter $C_{\rm Cas}=(2.34\pm 0.34)\cdot 
10^{-28}$ Hz${}^2$ m${}^{5}$, its sign confirming the expectation 
for an attractive force.  A final check on the parallelism between the 
surfaces is done by fitting the data with a function taking into 
account a deviation from the plane parallel geometry \cite{Bordag1}.
The resulting deviation  is $0 \pm 30$ nm, in such a way that no 
change is observed for $C_{\rm Cas}$ - this confirming the value 
estimated in the parallelization procedure.
From the definitions of $C_{\rm el}$ and $C_{\rm Cas}$ the coefficient 
of the Casimir force can be expressed as:
\vspace{-.2cm}
\begin{equation}
K_{C}= {\epsilon_{0} \over 4} 
{C_{\rm Cas} \over {C_{\rm el}}} = 
(1.22 \pm 0.18)\cdot 10^{-27} {\rm N m}^{2}.
\end{equation}
The value obtained agrees with the expected value of $K_C$ first evaluated 
by Casimir \cite{Casimir}. It can be noted from Fig. \ref{casimir} that 
the fitting curve does not describe systematically the experimental points 
in the 1-2 $\mu$m region. There are in principle several conventional 
effects which could be invoked to explain the observed deviation 
(see also Fig. \ref{casimir} caption) - such as border effects, 
residual roughness of the surfaces, finite conductivity of chromium 
or finite temperature corrections \cite{Bordag,Lambrecht}.  
Work is in progress to refine the data analysis handling the Casimir term 
at a level necessary to further subtract its contribution from the data. 
The control of Casimir force at this level and the evaluation of the 
residuals is necessary to test predictions, based on unification models 
of gravity to the electroweak and strong interactions, on new forces with 
intensity close to the gravitational force and acting below the millimeter 
range \cite{Fischbach,Antoniadis,Arkani}. 
The parallel plate configuration maximizes the sensitivity to such forces 
\cite{Carugno,Long}, and could lead to stronger constraints than the one 
already evaluated from Casimir forces in the plane-sphere configuration 
\cite{Bordag}. 

Our experimental verification of the Casimir prediction for the force between 
two parallel conducting surfaces in the 0.5 - 3.0 $\mu$m range leads to a 
measurement of the related coefficient with a 15$\%$ precision. 
Our results unambiguously show the existence of the quantum fluctuations at 
the macroscopic level, and confirm the existence of a delicate issue in 
matching quantum physics and the large scale evolution of the Universe via 
the cosmological constant problem. 
Furthermore, the technique demonstrated here and future refinements in 
various directions \cite{Bressi} could pave the road to a high precision 
control of the Casimir forces  crucial for both looking at new 
physics related to gravitation in the submillimeter range, as well as for  
designing electromechanical devices at the nanoscale. 
\begin{acknowledgments}
We thank Z. Fontana for initial contributions to the experiment,
F. Donadello, A. Galvani, and F. Veronese for skilful technical support, and
L. Viola for a critical reading of the manuscript.
The experiment was performed at the Laboratori Nazionali di Legnaro of 
the Istituto Nazionale di Fisica Nucleare.
\end{acknowledgments}

E-mail address: ruoso@lnl.infn.it

\newpage
Figure Captions

\begin{figure}[htb]
\caption{\label{set-up}
Experimental set-up. From left to right: displacement transducer, 
cantilever and opposing surface (source) solidal to a PZT actuator, 
capacitance meter and precision voltage source.
The two opposing surfaces, on which the Casimir force is studied, form
a capacitor with an area of 1.2 $\times$ 1.2 mm$^{2}$. 
The source, the PZT, its support and the motors are mechanically decoupled 
from the resonator by means of a set of alternated rubber rings and stainless 
steel disks. 
The apparatus is placed on the flange accessing the science chamber of
a Scanning Electron Microscope (SEM) Philips PSEM 500.  
This arrangement allows to perform distance measurement and control the 
gap with a resolution up to $50 \div 100$ nm, as well as to work at a 
residual pressure of $\sim 10^{-5}$ mbar, low enough to prevent direct 
acoustical pick-up and to mitigate both the formation of oxide on metallic 
surfaces and the relocation on the gap of ambient dust present in the SEM.}
\end{figure}

\begin{figure}[htb]
\caption{\label{sempicture}
Picture of the apparatus taken at the SEM. 
From top to bottom: source, cantilever, detection fiber. 
The fiber is covered by a grounded conducting cylinder to prevent charging 
from the electrons emitted from the SEM, the latter being turned off during 
the data acquisition. During the measurement the fiber end is located within 
20 to 50 $\mu$m from the resonator. This distance range provides a trade-off 
between optimizing the visibility in the interference signal and avoiding a 
direct influence of the fiber end on the resonator free frequency, e.g. due 
to residual charging. The field of view is 3 mm $\times$ 2.3 mm, and the gap 
is $d=110\,\,\mu$m.}
\end{figure}

\begin{figure}[htb]
\caption{\label{calibration}
Calibration with controllable electrostatic fields.
The square frequency difference is shown versus distance for the three 
different values of the bias voltage $V_c$, as well as the fits with the 
electrostatic function (Eq. (2)). 
Each frequency shift is evaluated through the FFT analyzer spectrum 
of the fiber interferometer signal (acquiring  two rms averages with 
resolution bandwidth of 31.25 mHz), which is then downloaded on a computer 
and fitted to a Lorenzian resonance curve to determine frequency and 
linewidth. The associated error arises from the Lorenzian fit 
plus a statistical uncertainty of 7 mHz.
The overall acquisition time is kept below 40 minutes  
to minimize the drifts in the system, e.g. of thermal origin.}
\end{figure}

\end{document}